\documentclass[AMA,Times1COL]{WileyNJDv5} %STIX1COL,STIX2COL,STIXSMALL

\articletype{Article Type}%

\received{00/00/0000}
\revised{Date Month Year}
\accepted{Date Month Year}
\volume{00}
\copyyear{2025}
\startpage{1}

\newcommand{\1}{\textbf{1}}
\def\1{1\!{\rm l}}

\newcommand{\bdz}{\boldsymbol z}
\newcommand{\bbeta}{\boldsymbol \beta}

\definecolor{blendedblue}{rgb}{0.2,0.2,0.7}
\definecolor{darkpurple}{RGB}{49,0,94}
\definecolor{darkgreen}{RGB}{11, 84, 37}
\usepackage{hyperref}

\newcommand{\given}{\,|\,}

\allowdisplaybreaks

\begin{document}

\title{Bayesian ensemble learning method for the analysis of multipollutant mixtures data}

\author[1]{Yu-Chien Ning}

%\author[2,3]{Donna Spiegelman}

\author[2]{Xin Zhou}

\author[4]{Francine Laden}

\author[1,5,6]{Molin Wang}

\authormark{Ning \textsc{et al.}}
\titlemark{Bayesian ensemble learning for predicting health outcomes of multipollutant mixtures}

\address[1]{\orgdiv{Department of Epidemiology}, \orgname{Harvard T.H. Chan School of Public Health}, \orgaddress{\state{MA}, \country{USA}}}

\address[2]{\orgdiv{Department of Biostatistics}, \orgname{Yale School of Public Health}, \orgaddress{\state{CT}, \country{USA}}}

\address[4]{\orgdiv{Departments of Environmental Health and Epidemiology}, \orgname{Harvard T.H. Chan School of Public Health}, \orgaddress{\state{MA}, \country{USA}}}

\address[5]{\orgdiv{Departments of Biostatistics}, \orgname{Harvard T.H. Chan School of Public Health}, \orgaddress{\state{MA}, \country{USA}}}

\address[6]{\orgdiv{Channing Division of Network Medicine}, \orgname{ Brigham at Women's Hospital and Harvard Medical School}, \orgaddress{\state{MA}, \country{USA}}}

\corres{Corresponding author Yu-Chien Ning, This is sample corresponding address. \email{bycning@hsph.harvard.edu}}

\presentaddress{677 Huntington Ave, Boston, MA 02115}

%\fundingInfo{Text}
%\JELinfo{ejlje}

\abstract[Abstract]{We introduce the SoftBart approach \citep{linero18} from Bayesian ensemble learning to estimate the relationship between multipollutant mixtures and health on chronic exposures in epidemiology research.
This approach offers several key advantages over existing methods: (1) it is computationally efficient and well-suited for analyzing large datasets; (2) it is flexible in estimating various correlated nonlinear functions simultaneously; and (3) it accurately identifies active variables within highly correlated multipollutant mixtures.
Through simulations, we demonstrate the method's superiority by comparing its accuracy in estimating and quantifying uncertainties for both main and interaction effects with the commonly used method, BKMR. Last, we apply the method to analyze a multipollutant dataset with 10,110 participates from the Nurses' Health Study.}


\jnlcitation{\cname{%
\author{Yu-Chien N},
\author{Donna S},
\author{Francine L}, and
\author{Molin W}}.
\ctitle{Bayesian ensemble learning method for the analysis of multi-pollutant mixtures} \cjournal{\it Statistics in Medicine} \cvol{2025;00(00):1--13}.}

\maketitle

%\renewcommand\thefootnote{}
%\footnotetext{\textbf{Abbreviations:} ANA, anti-nuclear antibodies; APC, antigen-presenting cells; IRF, interferon regulatory factor.}

\renewcommand\thefootnote{\fnsymbol{footnote}}
\setcounter{footnote}{1}

% INTRODUCTION
\section{Introduction} 
\label{sec:intro}

Estimating the relationship between individual health and air pollution is crucial, as it allows policymakers to develop effective public health strategies, monitor air pollutants, and improve overall health outcomes.
There are two major challenges in estimating this relationship: 1) the effects of air pollutants on health are often gradual---especially in areas less prone to severe air pollution---making the association between air pollution and health outcomes detectable often only in large datasets collected over long-term follow-up periods. In our paper, we will study a dataset consisting more than 10,000 observations that are collected from 1995 to 2008; and 2)  correlations among various multipollutant constituents can be strong and nonlinear, making it challenging to identify the dominant mixture components that have the most detrimental effects on health.
Despite significant progress has been made, there is a lack of ready-to-use statistical methods  that are both {\it computationally efficient}  for large-scale datasets and {\it flexible} enough to capture these interactions while identifying the specific pollutant components most strongly associated with adverse health outcomes.

In the past, multiple approaches have been proposed to tackle these two challenges. Popular methods include additive main effects models \citep{gold99, andrea14}, hierarchical models \citep{suh11}, and dimension reduction techniques, which transform multipollutant exposures into independent sets of variables and then examine their relationships with health outcomes using unsupervised methods \citep{ito05}. For a comprehensive review of this topic, see Davalos et al. (2017)\cite{davalos17}. However, these approaches either ignore higher-order interactions among exposures or rely on model parameters that are implicit and difficult to interpret. More importantly, they lack the ability to perform variable selection and estimate the effects of active components in a  nonparametric manner.

Recently, Bayesian machine learning methods have gained unprecedented popularity. One prominent example is the Bayesian Kernel Machine Regression (BKMR) method proposed by Bobb et al. (2015) \cite{bobb15}. This approach has proven successful in capturing nonlinear correlations within high-dimensional multipollutant exposures and, at the same time, performing variable selection to identify dominant components that significantly impact individual health outcomes. The BKMR method allows users to easily obtain posterior means, medians, and various uncertainty quantifications through Markov chain Monte Carlo (MCMC) simulations. In its core, BKMR employs a Gaussian process prior for curve fitting and a spike-and-slab prior for model selection---both of which have been extensively studied in recent statistical literature \citep{vdV08, castillo15, ning20, jiang21}. However, a significant limitation for practical use is that it is extremely slow computationally and can become infeasible for analyzing large datasets for a reason as we will explain below.

In this paper, we introduce the SoftBart method, originally proposed by Linero \& Yang (2018)\cite{linero18}, for the analysis of multipollutant data. This method is a Bayesian ensemble learning approach that models nonlinear relationships using a combination of ``soft'' trees (see Model \ref{tree} for details) and performs variable selection through a soft thresholding procedure. Compared to the spike-and-slab prior in BKMR, which relies on hard thresholding, this soft thresholding approach can significantly improve computational speed, making SoftBart a competitive alternative to BKMR for analyzing large datasets.

\begin{figure}[!h]
\centering
\includegraphics[width=0.5\textwidth]{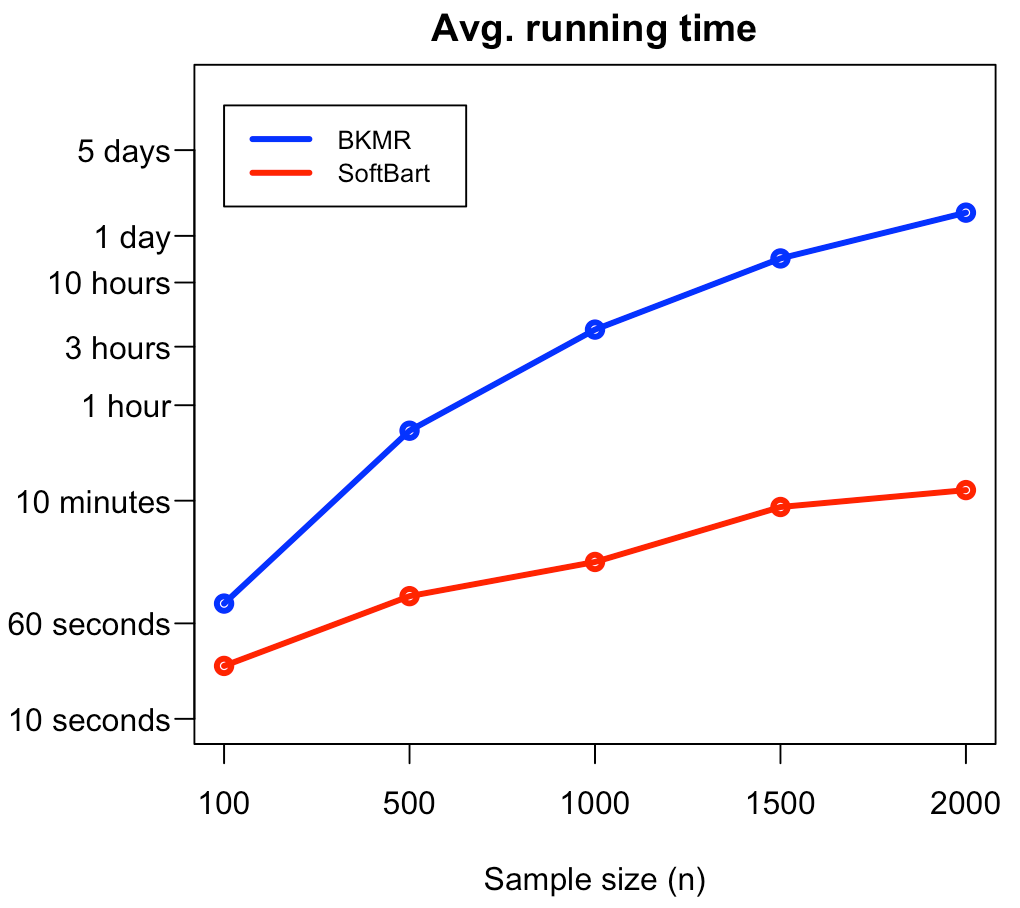}
\vspace{-0.2cm}
\caption{Average running CPU time for the BKMR and SoftBart methods; time is displayed on a log scale.}
\label{fig:runtime}
\end{figure}

To compare the computational speed of SoftBart and BKMR, we conducted a simulation study (details provided in Section \ref{sec:sim}) and found that SoftBart is significantly faster. For a dataset with a sample size of 100, SoftBart is three times as fast as BKMR, and for a sample size of 2000, it is 182 times faster---BKMR takes about 37 hours (CPU time), whereas SoftBart completes the computation with the same number of MCMC iterations in just 12 minutes.  
In Figure \ref{fig:runtime}, we plot the computation time of the two methods with sample sizes from 100 to 2000. Due to limitations in our computational resources, we were unable to report the running time for sample sizes larger than 2000 for BKMR.
A key reason for the high computation cost of BKMR for even a moderate size dataset is the use of the spike-and-slab prior for variable selection. This approach requires the algorithm evaluating $2^M$ combinations among $M$ components in each MCMC iteration. The SoftBart method, on the other hand, uses the Dirichlet distribution for variable selection, which avoids this combinatorial evaluation, and thus significantly reduces computation time.

\begin{figure}[!h]
\centering
\begin{minipage}{0.47\textwidth}
\centering
\includegraphics[width=1\textwidth]{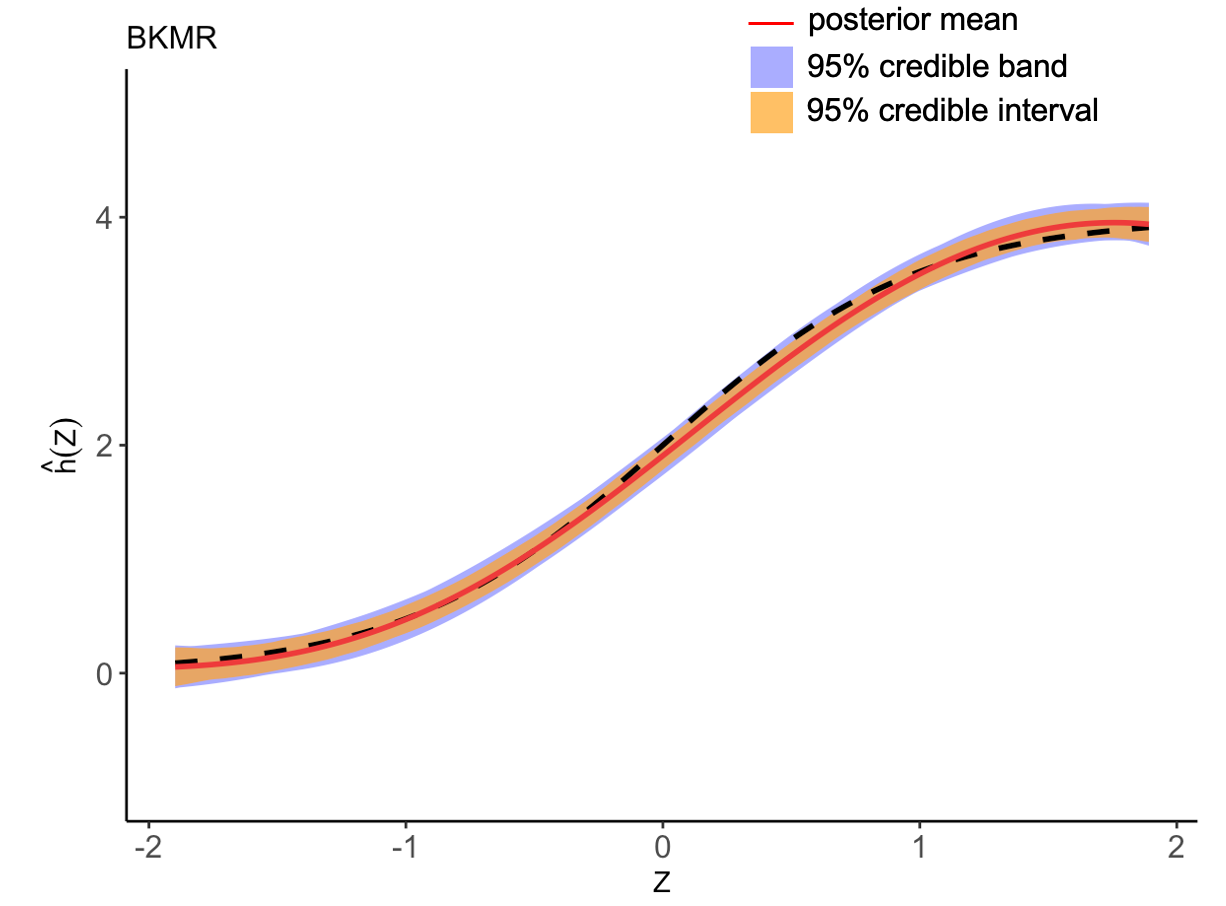}
\end{minipage}%
\hspace{0.04\textwidth}
\begin{minipage}{0.47\textwidth}
\includegraphics[width = 1\textwidth]{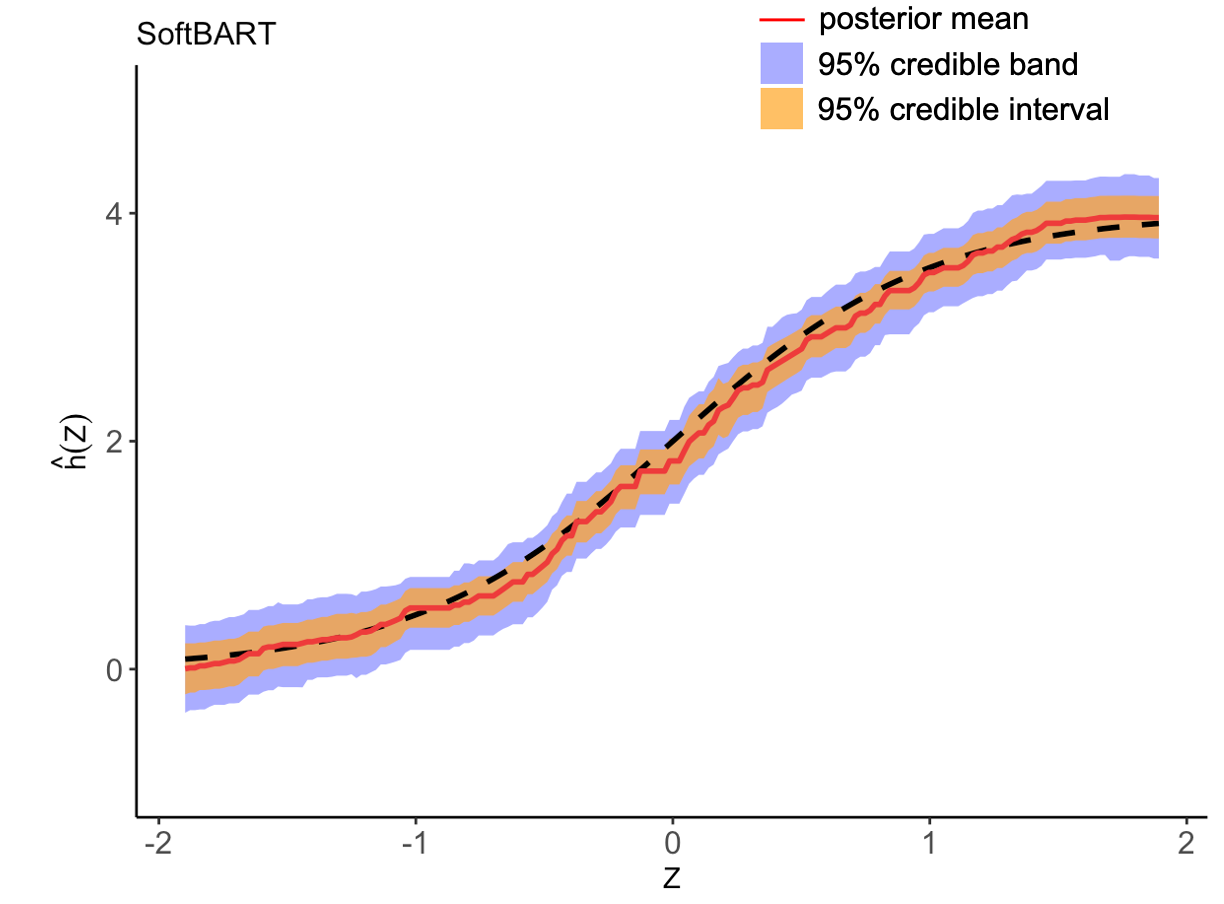}
\end{minipage}
\begin{minipage}{0.47\textwidth}
\centering
\includegraphics[width=1\textwidth]{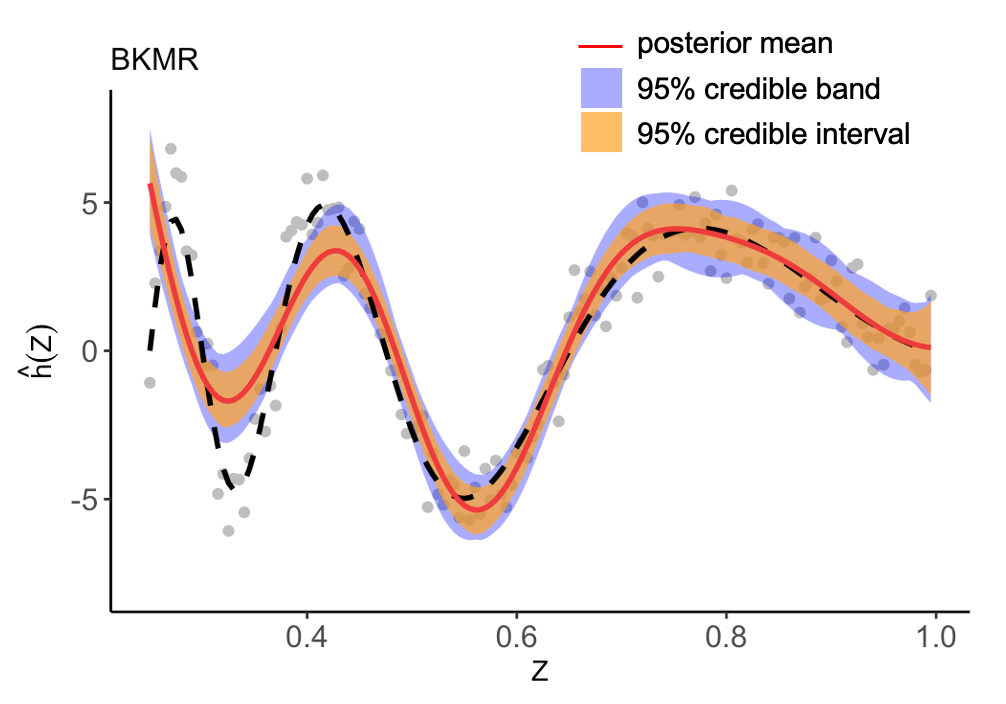}
\end{minipage}%
\hspace{0.04\textwidth}
\begin{minipage}{0.47\textwidth}
\includegraphics[width = 1\textwidth]{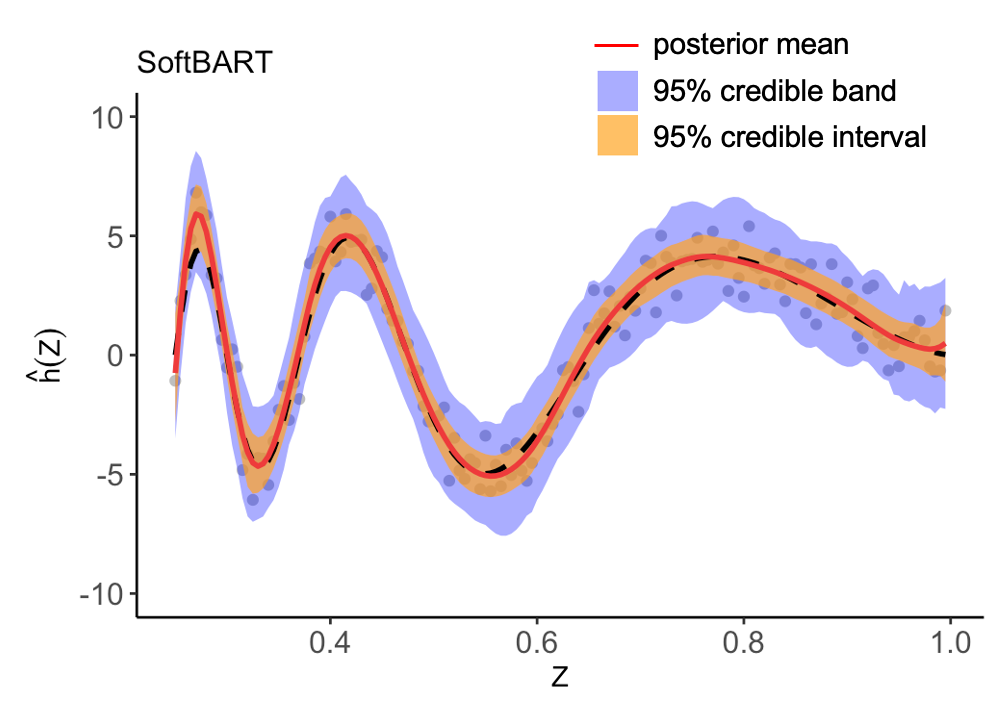}
\end{minipage}
\vspace{-.4cm}
\caption{Posterior means (red) and 95\% credible intervals (orange) and credible bands (blue) of a function with a varying smoothness estimated by BKMR (left) and the SoftBart (right). Unlike SoftBart, BKMR has a large bias in estimating the curve.}
\label{fig:nsmooth-fn}
\end{figure}

Besides its computational advantages, SoftBart has another advantage over BKMR. Although both methods are capable of capturing various nonlinear relationships, SoftBart can estimate functions in spaces that extend beyond the Reproducing Kernel Hilbert Space (RKHS). This capability allows SoftBart to estimate functions with varying levels of smoothness.
To illustrate this, we generate two functions, one belonging to the RKHS and the other not, as shown in Figures \ref{fig:nsmooth-fn} (see the black line in both plots). 
Data were generated based on each function, and both BKMR and SoftBart were applied to fit the data. Since the functions are one-dimensional, no variable selection was performed. Details of the data generation process are discussed in Section \ref{sec:sim}.
From the results in the two figures, we observe that while BKMR fails to estimate the function in Figure \ref{fig:nsmooth-fn}, SoftBart performs well for both functions. Note that the underwhelming result of BKMR cannot be addressed by hyperparameter tuning. In practice, because the underlying function space is often unknown, it is advantageous to use a method as general and flexible as SoftBart.
In sum, while both BKMR and SoftBart are Bayesian nonparametric methods that are flexible in estimation and capable of providing uncertainty quantification for unknown functions, SoftBart has advantages over BKMR for estimating large datasets.

The remaining parts of the paper is organized as follows: In Section \ref{sec:model} we introduce the model and both BKMR and SoftBart methods; Section \ref{sec:sim} includes multiple simulation studies to exam the performance of the two methods for estimating various functions. 
In Section \ref{sec:real}, we apply the SoftBart method to estimate the associations of 12 multipollutant constituents of PM2.5 with the cognitive function in the Nurses' Health Study (NHS), which includes 10,110 participants.
Last, in Section \ref{sec:disc}, we conclude the paper and discuss key future research directions. 

%%%%%%%%%%%%%%%%%%%%%%%%%%%%%%
\section{The Bayesian partial linear model}
\label{sec:model}

Consider the dataset containing observations $\mathcal{D}^n = {(Y_i, Z_i, X_i)}_{i=1}^n$ from $n$ i.i.d. individuals. For the $i$-th individual, let $Y_i \in \mathbb{R}$ represent the health endpoint, $Z_i \in \mathbb{R}^M$ a vector containing $M$ multipollution constituents, and $X_i \in \mathbb{R}^d$ a set of potential confounders.
The partial linear model is given as follows: 
\begin{align} 
\label{model} 
Y_i = h(Z_i) + X_i' \bbeta + \varepsilon_i, \ \varepsilon_i \overset{\mathrm{i.i.d}}{\sim} N(0, \sigma^2), 
\end{align} 
where $\bbeta \in \mathbb{R}^d$ is the vector of regression coefficients, $\sigma^2 \in \mathbb{R}^+$ is the error variance, and $h \in L^2[0, 1]^M$ is a function characterizing the relationship between  the $M$ variables and the outcome.  
This function is allowed to possess the following properties: 1) sparsity, which is the key property to allow selecting a subset of components that are significantly correlated with the health endpoint, and 2) nonlinearity, where both main and interaction effects, and even higher order interaction effects can be modeled nonlinearly. 

We consider two different nonparametric priors for modeling $h(\cdot)$, where BKMR uses the sparse Gaussian processes prior and SoftBart is uses ensemble of `soft' trees. Besides $h(\cdot)$, we consider priors for common parameters in Model \ref{model} as follows: the noninformative prior for $\beta \propto 1$ and the inverse Gamma prior for $\sigma^2 \sim \text{IG}(a, b)$. 

\subsection{The BKMR prior}

The BKMR prior for $h(\bdz)$ is given as follows:
\begin{align} 
\label{bkmr-prior} 
h(\bdz) \sim \mathcal{GP}(0, K(\bdz, \bdz')), \quad K(\bdz, \bdz') := K(\bdz, \bdz'; \boldsymbol r, \rho) = \rho^2 e^{ - \sum_{m=1}^M r_m (z_m - z_m')^2}, 
\end{align} 
where $K(\cdot, \cdot)$ is the squared exponential kernel, $\rho^2 \in \mathbb{R}^+$ is the overall variance of the kernel, and $\boldsymbol r = (r_1, \dots, r_M)$, with each $r_m$ as the inverse length scale for the corresponding $m$-th variable.

If variable selection is not needed, one can simply set $r_1, \dots, r_M = r \in \mathbb{R}^+$. Otherwise, a spike-and-slab prior is placed on $\boldsymbol r$, i.e., for each $j \in {1, \dots, M}$, let $\pi(\boldsymbol r) = \bigotimes_{j=1}^M \pi(r_j)$,  
\begin{align} \label{sas} 
r_m \given \gamma_m \sim \gamma_m f_1(r_m) + (1-\gamma_m) \delta_0, \quad \gamma_m \given \theta \sim \text{Bernoulli}(\theta), 
\end{align} 
where $f_1(r_m)$ as a probability density on $\mathbb{R}^+$, $\delta_0$ as the Dirac measure at 0, and $\gamma_m \in \{0, 1\}$ for all $m \in \{1, \dots, M\}$. 
If $\gamma_m = 1$, the $m$-th variable is included in the model. Conversely, if $\gamma_m = 0$, it is excluded, and the corresponding main effect of $z_m$ is modeled as a constant. The hyperparameter $\theta$ can be set to a fixed value between 0 and 1, though it is generally preferable to put a Beta prior, e.g., $\text{Beta}(1, M)$, on $\theta$ for theoretical considerations\cite{rockova15}.
The parameter $\rho$ in \eqref{bkmr-prior} can be estimated either via an empirical Bayes approach by maximizing the marginal posterior distribution or an hierarchical Bayes approach estimated via the MCMC algorithm as in the BKMR approach\cite{bobb15}.

\subsection{The SoftBart prior}
The SoftBart prior method models $h(\cdot)$ as a linear combination of tree priors given by 
\begin{align*} 
h(\bdz) = \sum_{t =1}^T \text{Tree} (\bdz; \mathcal{T}_t, \mathcal{M}_t), 
\end{align*}
where $\mathcal{T}_t$ is the $t$-th decision tree and $\mathcal{M}_t$ is the corresponding set of leaf node parameters.
Each tree consists of a collection of branches specified by 
\begin{align} \label{tree} 
\text{Tree}(\bdz; \mathcal{T}, \mathcal{M}) = \sum_{ \ell \in \mathcal{L}(\mathcal{T})} \mu_{\ell} \phi_\ell(\bdz; \mathcal{T}), 
\end{align} 
where $\mathcal{L}(\mathcal{T})$ represents the set of leaf nodes in $\mathcal{T}$, and $\mu_\ell$ is the coefficient for each basis function $\phi_\ell(\cdot)$, which constructs the tree.
Each tree is referred to as a "soft" tree, as $\phi_\ell$ is a continuous and differentiable function, which is given by 
\begin{align} \label{soft-rule} 
\phi_\ell(\bdz; \mathcal{T}) = \prod_{b \in \mathcal{A}(\ell)} \psi \left(\frac{\bdz_{j_b} - C_b}{\tau} \right)^{L_b} \bar\psi \left(\frac{\bdz_{j_b} - C_b}{\tau} \right)^{1-L_b},
\end{align} 
where $\mathcal{A}(\ell)$ represents the set of nodes ancestral to the leaf node $\ell$, $\psi(k) = {1}/(1+e^{-k})$ is the logistic function, 
and $\bar \psi(\cdot) = 1 - \psi(\cdot)$. The parameter $\tau$ is the bandwidth and $L_b$ is a binary variable indicating the path direction from the root (i.e., $L_b = 1$ for left, and $L_b = 0$ otherwise). The cut point $C_b$ in \eqref{soft-rule} specifies the splitting rule in each branch as ${\bdz_{j_b} \leq C_b}$.
The trees in the classical BART model \cite{chipman10} can be understood as hard trees, since each basis function is an indicator function, which corresponds to the function in \eqref{soft-rule} with $\tau \to 0$.

Compared to the BKMR prior in \eqref{bkmr-prior}, the dual form of the Gaussian process representation allows $h(\bdz)$ to be expressed as a sum of basis functions: $h(\bdz) = \sum_{l=1}^L a_l \phi_l(z)$, where $\phi_l(\cdot)$ denotes a basis function such that the inner product $\langle \phi_l, \phi_k \rangle$ equals 0 if $l \neq k$ and 1 if $l = k$, and $a_l$ is the unknown coefficient to be estimated. The SoftBart prior \eqref{soft-rule} differs fundamentally from the BKMR prior in its choice of basis functions.

Parameters of the SoftBart prior can be categorized into two sets: the first set characterizes the tree shape, with depth $d$ assigned a probability $\gamma (1 + d)^{-\beta}$ ($\gamma, \beta > 0$) to limit excessive tree growth. 
The second set determines the splitting rule for each branch's cut point, chosen by first sampling a predictor as $j_b \sim \text{Categorical}(s)$, where $s = (s_1, \dots, s_p)$, and then sampling the cut point $C_b$ from a uniform distribution. We refer Linero \& Yang (2018) \cite{linero18} to the details of the prior selection. 

Variable selection in SoftBart is performed by placing a Dirichlet prior on $s \sim \text{Dir}(a/p, \dots, a/p)$, with $a/(a + p) \sim \text{Beta}(0.5, 1)$. Notably, sampling from the Dirichlet prior is faster than the spike-and-slab prior \eqref{sas}, which is frequently used in Bayesian variable selection, as it avoids evaluating the $2^M$ inclusion probabilities of $M$ variables at each MCMC iteration. 
Priors for the remaining hyperparameters in \eqref{soft-rule} are chosen as follows: 
$\tau \sim \exp(0.1)$, $C_b \sim \text{Uniform}(a, b)$ where $a$ and $b$ are the starting and ending points of interval determined by the path to $b \in \mathcal{A}(\ell)$.

Last, we set $\beta = 2$ in the prior of the tree shape and choose $\gamma = 0.3$ for a sample size of $n = 100$ and $\gamma = 0.1$ for a sample size of $n = 1000$. We found that if choosing the value $\gamma$ as the default value in \citep{linero22} (which is 1), then the method would not penalize the depth of the trees enough, the method does not sufficiently penalize the depth of the trees, leading to excessively large credible sets for the estimated value of $f(\bdz)$.

\section{Simulation study}
\label{sec:sim}

We conduct simulation studies to evaluate the performance of the SoftBart method, in comparison with the BKMR method. In each simulation, data are generated based on Model \eqref{model} as follows: we first specify $h(\boldsymbol{z})$; then set the sample size $n$ and the number of mixture components $M$. Next, we fix $\beta_1 = \beta_2 = 2$ and generate two synthetic confounder variables as follows: $X_{1i} \overset{iid}{\sim} N(0, 1)$ and $X_{2i} \overset{iid}{\sim} \cos(Z_{1i}) + N(0, 0.2)$. Finally, we generate 100 datasets with the residual variance $\sigma^2 = 0.5$. For each dataset, we apply both methods separately. Throughout our simulations, we use two sample sizes, $n = 100$ and $n = 1000$, and consider two values for $M$, $M = 3$ and $M = 13$.

We consider four types of $h(\cdot)$, each characterized by a unique combination of sparsity and smoothness levels, allowing us to explore a range of functional behaviors and their impact on model performance; they are
\begin{enumerate}
\item $\bdz \in \mathbb{R}^3$, $h^{(1)}(\bdz) = 2 \sin( \pi z_1/2) + 2\cos(\pi z_2/2) + 
    (z_1 + z_2)^2$;
\item $\bdz \in \mathbb{R}^3$, $h^{(2)}(\bdz) = h^{(1)}(\bdz)  + 4 \times \text{plogis}(z_3; 0, 0.3)$,
where $\text{plogis}(x; a, b) = (1+\tanh((x - a)/(2b))/2$ and $\tanh(x)$ is the hyperbolic tangent of $x$;
%\item $\bdz \in \mathbb{R}^3$, $h^{(3)}(\bdz) = 4 (z_1 + z_2) + 2 z_1  z_2$;
\item $\bdz \in (0.2, 1] \times \mathbb{R}^2$, 
$h^{(3)}(\bdz) =  10 \times \sqrt{z_1 (1-z_1)} \sin(2\pi /z_1)
+ 4 \times \text{plogis}(z_2; 0, 0.3) + (z_1 + z_2)^2$;
\item $\bdz \in \mathbb{R}^{13}, h^{(4)}(\bdz) = h^{(1)}(\bdz)$.
\end{enumerate}
The first three functions involve three correlated variables with correlation coefficients $\text{corr}(z_1, z_2) = 0.1$, $\text{corr}(z_2, z_3) = 0.3$, and $\text{corr}(z_1, z_3) = 0.7$. Among these, the first and third functions exclude $z_3$, with the distinction between them being the form $h^{(3)}(z_1)$. In the third function, $h^{(3)}(z_1)$ exhibits varying smoothness with respect to $z_1$, where SoftBart is expected to outperform in estimating this function. All three variables are included in the second function.
By running simulations on the three functions will allow us to evaluate the variable selection performance of the two methods under both sparse and dense scenarios.

The last function involves 13 variables, with only the first two included as active variables in the model. The correlation structure among 13 variables is derived from the air pollution dataset, which contains daily measurements at a central site in Boston from 1999 to 2011, with the correlation matrix provided in the BKMR package \citep{bkmr}. This setup allows us to test the methods in a scenario that closely mimics real-world data. The simulation results for the first three functions are given in Section \ref{sec:sim-m3} and for the last function are provided in Section \ref{sec:sim-m13}.

\subsection{Simulations with three mixture components}
\label{sec:sim-m3}
In this section, we compare the two methods for the choice of $h(\cdot) = h^{(1)}(\cdot)$, $h^{(2)}(\cdot)$, or $h^{(3)}(\cdot)$ respectively. We ran BKMR using the `bkmr' package \citep{bkmr} in $\mathsf{R}$ and 
choose the prior for $\rho$ and $r_m$ in \eqref{bkmr-prior} the same as the default priors in the BKMR package as follows: $\rho = \sigma^2 \lambda$ with $\lambda$ follows a gamma ditribution with mean and variables are both 100 and $r_m \sim \text{Unif}(0, 100)$.
For SoftBart, we used the `SoftBart' package in $\mathsf{R}$ and set the number of trees to 30. As we discussed in the last paragraph of Section \ref{sec:model}, tuning $\gamma$ is also critical since it regulates tree depth. Larger $\gamma$ values can result in overly wide credible sets, including credible intervals and credible bands. For sample size $n = 100$, we set $\gamma = 0.3$, and for $n = 1,000$, we chose $\gamma = 0.1$.

Both methods adopt the MCMC algorithm to compute their posterior distributions. For each method, we generated 20,000 total draws and evaluated the following metrics: (1) regression coefficients and $R^2$ values obtained by regressing the main effects $h(z_m)$ that are used to generate the data on their posterior means $\hat{h}(z_m)$ for $m = 1, 2, 3$; (2) lengths of 95\% pointwise credible intervals and 95\% credible bands\footnote{We consider fixed width simultaneous credible bands centered at the posterior mean $\hat f$ at level $\gamma \in (0, 1)$ denoted by $C(r) = \{f: \|f - \hat f\|_\infty \leq r\}$, where $\|\cdot\|_\infty$ is the $L_\infty$-norm of functions. The value $r$ is the half length so that the posterior probability of $f$ falling into the credible band $C(r)$ equal to $r$, i.e., $\Pi(f: f \in C(r) \given X_n) = r$.} for the posterior of $h(z_m)$, along with coverage probabilities of the true values; (3) posterior inclusion probabilities (PIPs) for variable selection; and (4) the average runtime of each algorithm. Post-MCMC computation times for quantities described in (1)--(3) are excluded. Simulations were performed in the Cannon cluster at FASRC at Harvard\footnote{Cannon cluster: \url{https://www.rc.fas.harvard.edu/about/cluster-architecture/}}.

In our simulations, we computed 95\% pointwise credible intervals as well as 95\% credible bands for each function $f$. The credible bands provide a more comprehensive evaluation of the methods' performance across the entire estimated function, while the piecewise credible intervals that are calculated at each given point of $f$ only guarantees posterior coverage at that point. Therefore, in general, the 95\% piecewise credible intervals are shorter than the 95\% credible bands.
We also evaluated the two-way interaction effects of multipollutant components.
For BKMR, the main and interaction effects can be obtained directly from the marginal posterior distributions, which are Gaussian processes, as the joint posterior is a multivariate Gaussian process. In contrast, estimating these effects for SoftBart is less straightforward. It requires evaluating the partial dependence function \citep{friedman01} by numerically integrating over the remaining variables in the joint posterior distribution.

\begin{table}[!h]
\centering
\caption{Main effects estimated from the SoftBart and BKMR methods ($n = 100$). We report the intercept, slope, and $R^2$ for the regression of $h(\cdot)$ of $\hat h(\cdot)$, the lengths of the 95\% piecewise credible interval and the 95\% credible band, the coverage probability of the credible band, and the posterior inclusion probability (PIP).}
\label{tab:sim-main-n100}
\begin{tabular}{c c | c c c | c c c | c}
\toprule
& & \multicolumn{3}{c|}{Regression of $h(\cdot)$} & \multicolumn{3}{c|}{Uncertainty Quantification ($\alpha$ = 0.95)} &  \\
Main effect & Method & Intercept & Slope & $R^2$ & len(C.I.) & len(C.B.)  & Coverage & PIP\\
\midrule
\multicolumn{9}{c}{Case 1: $h^{(1)}(\boldsymbol z)$}\\
\toprule
$h(z_1)$  & SoftBart &  0.11 &  0.96 & 0.95  & 1.73 & 2.95 & 0.88 & 1.00  \\
               & BKMR       & 0.07 & 0.98  & 0.99  & 1.48 &  2.22 & 0.98  &  1.00  \\ 
$h(z_2)$  &  SoftBart & 0.79 &  0.71 &  0.51 &  1.67 & 2.82 & 0.89  & 0.99  \\  
               & BKMR        & 0.51 &  0.60 & 0.53  & 1.40 &  2.10 & 0.86 & 1.00  \\ 
$h(z_3)$  &  SoftBart &  $3.79$ & $-0.18$ & $0.02$ & $1.16$ & $1.18$  & 1.00 & 0.11  \\
               & BKMR        &  3.41 & $-1.00$ & $0.03$ & 1.37 & 1.54 & 1.00 & 0.10 \\
\midrule                               \\[-6pt]
\multicolumn{2}{c}{Avg. running time} & \multicolumn{6}{c}{SoftBart: 27.46s, \quad BKMR: 87.4s} & \\[2pt]
\toprule
\multicolumn{9}{c}{Case 2: $h^{(2)}(\boldsymbol z)$}\\
\midrule
$h(z_1)$ & SoftBart &  0.09  & 0.98 &  0.97 &  1.53 & 2.99 &  0.96 & 1.00 \\ 
				& BKMR       & $-0.07$ & 1.01 &  0.96 & 1.82 & 2.75 & 0.93 & 1.00\\
$h(z_2)$ &  SoftBart & 1.20  & 0.74  &  0.54 &  1.37 &  2.63 & 0.90 & 1.00 \\  
			   & BKMR        & $2.21$  & 0.32 &  0.24 & 1.57 & 2.28 & 0.72 & 1.00 \\
$h(z_3)$ & SoftBart & 0.09 & 0.98 & 0.99 & 1.47 & 2.66 & 1.00 & 1.00 \\
				& BKMR &       $-0.49$  & 1.14 & 0.94 & 1.86 & 2.70 & 0.81 & 1.00\\
\midrule                               \\[-6pt]
\multicolumn{2}{c}{Avg. running time} & \multicolumn{6}{c}{SoftBart: 26.59s, \quad BKMR: 98.2s} & \\[2pt]
\toprule
\multicolumn{9}{c}{Case 3: $h^{(3)}(\boldsymbol z)$}\\
\midrule
$h(z_1)$ & SoftBart &  0.60 & 0.89 & 0.91 & 2.94 & 6.41 & 0.61 & 1.00 \\ 
			& BKMR          &  0.28 & 0.93 & 0.90 &  3.29 & 5.22 & 0.37 & 1.00 \\
$h(z_2)$ &  SoftBart &  0.18 & 0.96 & 0.97 & 1.91 & 3.60 & 0.98 & 1.00 \\  
				& BKMR       & 0.01 & $0.95$ & $0.95$ & 3.64 & 5.29 & 1.00 & 1.00 \\
$h(z_3)$ & SoftBart &  1.30 & $0.82$  & 0.02 & 1.33 & 1.67 & 1.00 & 0.06 \\
				& BKMR & $-1.18$ & $-0.65$ & 0.02 & 3.02  & 3.23 & 1.00 & 0.09  \\
\midrule                                  \\[-6pt]
\multicolumn{2}{c}{Avg. running time} & \multicolumn{6}{c}{SoftBart: 21.59s, \quad BKMR: 83.71s } & \\[2pt]
\bottomrule			    
\end{tabular}
\end{table}

\begin{table}
\centering
\caption{Interaction effects estimated from the SoftBart and BKMR methods ($n = 100$).
We report the intercept, slope, and $R^2$ for the regression of $h(\cdot)$ of $\hat h(\cdot)$, the lengths of the 95\% piecewise credible interval and the 95\% credible band, and the coverage probability of the credible band.}
\label{tab:sim-int-n100}
\begin{tabular}{c c | c c c | c c c }
\toprule
Interaction & & \multicolumn{3}{c|}{Regression of $h(\cdot)$} & \multicolumn{3}{c}{Uncertainty Quantification ($\alpha$ = 0.95)}  \\
effect & Method & Intercept & Slope & $R^2$ & len(C.I.) & len(C.B.)  & Coverage\\
\midrule
\multicolumn{8}{c}{Case 1: $h^{(1)}(\boldsymbol z)$}\\
\midrule
$h(z_1, z_2)$  & SoftBart &  0.15 &  0.94 & 0.97  & 2.67 & 5.35 & 0.92   \\
               & BKMR       	       & 0.10 & 0.96  & 0.98  & 1.67 &  2.74 & 0.33 \\ 
$h(z_1, z_3)$  &  SoftBart & 0.11 &  0.96 &  0.95 &  1.83 & 3.21 & 0.90    \\  
               & BKMR        & 0.09 &  0.97 & 0.99  & 1.57 &  2.41 & 0.98  \\ 
$h(z_2, z_3)$  &  SoftBart & 0.80 & 0.70 & 0.49 & 1.80 & 3.15 & 0.95  \\
               & BKMR        &  0.55 & $0.58$ & $0.49$ & 1.49 & 2.28 & 0.95 \\
\toprule
\multicolumn{8}{c}{Case 2: $h^{(2)}(\boldsymbol z)$}\\
\midrule
$h(z_1, z_2)$  & SoftBart &  0.25 &  0.95 & 0.98  & 2.32 & 5.09 & 0.95   \\
               & BKMR       	       & 0.25 & 0.95  & 0.96  & 2.12 &  3.39 & 0.24 \\ 
$h(z_1, z_3)$  &  SoftBart & 0.09 &  0.98 &  0.97 &  2.01 & 4.38 & 0.96    \\  
               & BKMR        & 0.14 &  0.97 & 0.92  & 2.41 &  3.79 & 0.74  \\ 
$h(z_2, z_3)$  &  SoftBart & 0.16 & 0.97 & 0.94 & 1.95 & 4.14 & 0.95  \\
               & BKMR        &  $-0.20$ & $1.06$ & $0.88$ & 2.13 & 3.32 & 0.43 \\
\toprule
\multicolumn{8}{c}{Case 3: $h^{(3)}(\boldsymbol z)$}\\
\midrule
$h(z_1, z_2)$  & SoftBart & 0.49 & 0.91  & 0.94  & 3.85 & 9.80 & 0.82   \\
               & BKMR       	       & 0.52 & 0.90  & 0.90  & 4.08 &  6.76 & 0.14 \\ 
$h(z_1, z_3)$  &  SoftBart & 0.63 & 0.88 & 0.91 & 3.04  & 6.83 & 0.78   \\  
               & BKMR        & 0.37 &  0.90 &  0.87  & 3.45 &  5.51 & 0.39 \\ 
$h(z_2, z_3)$  &  SoftBart &  0.18 & 0.96 & 0.97 & 2.01 & 4.05 & 0.99  \\
               & BKMR        & 0.01 & 0.94 & 0.94 & 3.76 & 5.56 & 1.00 \\
\bottomrule
\end{tabular}
\end{table}

\begin{table}
\centering
\caption{Main effects estimated from the SoftBart and the BKMR methods ($n = 1000$). We report the intercept, slope, and $R^2$ for the regression of $h(\cdot)$ of $\hat h(\cdot)$, the lengths of the 95\% piecewise credible interval and the 95\% credible band, the coverage probability of the credible band, and the posterior inclusion probability (PIP).}
\label{tab:sim-main-n1000}
\begin{tabular}{c c | c c c | c c c | c  }
\toprule
& & \multicolumn{3}{c|}{Regression of $h(\cdot)$} & \multicolumn{3}{c|}{Uncertainty Quantification ($\alpha$ = 0.95)} &   \\
Main effect & Method & Intercept & Slope & $R^2$ & len(C.I.) & len(C.B.)  & Coverage & PIP\\
\midrule
\multicolumn{9}{c}{Case 1: $h^{(1)}(\boldsymbol z)$}\\
\midrule
$h(z_1)$  & SoftBart &  $-0.01$ &  1.01 & 1.00  & 0.53 & 0.93 & 0.95 & 1.00  \\
               & BKMR       & 0.04 & 0.99  & 1.00  & 0.41 &  0.64 & 0.97  &  1.00  \\ 
$h(z_2)$  &  SoftBart & 0.16 &  0.94 &  0.92 &  0.54 & 0.96 & 0.89  & 1.00  \\  
               & BKMR        & 0.00 &  1.00 & 0.97  & 0.41 &  0.64 & 0.96 & 1.00  \\ 
$h(z_3)$  &  SoftBart &  $3.52$ & $-0.20$ & $0.33$ & 0.33 & 1.00  & 1.00 & 0.16  \\
               & BKMR        &  2.04 & 0.00 & $0.02$ & 0.36 & 0.36 & 1.00 & 0.00 \\
\midrule                                  \\[-6pt]
\multicolumn{2}{c}{Avg. running time} & \multicolumn{6}{c}{SoftBart: 180.62s, \quad BKMR: 14880.5s} & \\[2pt]
\toprule
\multicolumn{9}{c}{Case 2: $h^{(2)}(\boldsymbol z)$}\\
\midrule
$h(z_1)$ & SoftBart &  0.00  & 1.00 & 1.00 & 0.45 & 0.96 & 0.95 & 1.00 \\ 
				& BKMR     & 0.09 & 0.98 &  0.99 & 0.64 & 0.96 & 0.94 & 1.00\\
$h(z_2)$ &  SoftBart & 0.24  & 0.95  &  0.93 &  0.41 &  0.85 & 0.85 & 1.00 \\  
			   & BKMR      & $-0.03$  & 1.01 &  0.95 & 0.52 & 0.80 & 0.97 & 1.00 \\
$h(z_3)$ & SoftBart & 0.01 & 1.00 & 1.00 & 0.44 & 0.83 & 1.00 & 1.00 \\
				& BKMR &    0.08  & 0.98 & 0.99 & 0.79 & 1.28 & 0.32 & 1.00\\
\midrule                                  \\[-6pt]
\multicolumn{2}{c}{Avg. running time} & \multicolumn{6}{c}{SoftBart: 165.83s, \quad BKMR: 12787.0s} & \\[2pt]
\toprule
\multicolumn{9}{c}{Case 3: $h^{(3)}(\boldsymbol z)$}\\
\midrule
$h(z_1)$ & SoftBart &  0.21 & 0.96 & 0.99 & 1.62 & 4.36 & 0.68 & 1.00 \\ 
			& BKMR          &  $-0.03$ & 1.01 & 0.99 & 0.78 & 1.25 & 0.18 & 1.00 \\
$h(z_2)$ &  SoftBart &  0.07 & 0.99 & 1.00 & 0.72 & 1.48 & 0.85 & 1.00 \\  
				& BKMR       &  0.00 & 0.99 & $0.99$ & 0.87 & 1.32 & 0.90 & 1.00 \\
$h(z_3)$ & SoftBart &  5.41 & 0.01 & 0.01 & 0.39 & 0.60 & 1.00 & 0.09 \\
				& BKMR & $-1.30$ & 0.00 & 0.02 & 0.72  & 0.72 & 1.00 & 0.00  \\
\midrule                                   \\[-6pt]
\multicolumn{2}{c}{Avg. running time} & \multicolumn{6}{c}{SoftBart: 196.58s, \quad BKMR: 13232.6s } & \\[2pt]
\bottomrule					    
\end{tabular}
\end{table}

\begin{table}[h!]
\centering
\caption{Interaction effects estimated from the SoftBart and BKMR methods ($n = 1000$). We report the intercept, slope, and $R^2$ for the regression of $h(\cdot)$ of $\hat h(\cdot)$, the lengths of the 95\% piecewise credible interval and the 95\% credible band, and the coverage probability of the credible band.}
\label{tab:sim-int-n1000}
\begin{tabular}{c c | c c c | c c c }
\toprule
Interaction & & \multicolumn{3}{c|}{Regression of $h(\cdot)$} & \multicolumn{3}{c}{Uncertainty Quantification ($\alpha$ = 0.95)}  \\
effect & Method & Intercept & Slope & $R^2$ & len(C.I.) & len(C.B.)  & Coverage\\
\midrule
\multicolumn{8}{c}{Case 1: $h(\bdz) = h^{(1)}(\boldsymbol z)$}\\
\midrule
$h(z_1, z_2)$  & SoftBart &  0.03 &  0.99 & 1.00  & 0.95 & 1.95 & 0.96   \\
               & BKMR       	       & 0.02 & 0.99  & 1.00  & 0.49 & 0.80 & 0.42 \\ 
$h(z_1, z_3)$  &  SoftBart & $-0.01$ &  1.00 &  1.00 &  0.56 & 1.01 & 0.96    \\  
               & BKMR        & 0.05 &  0.99 & 1.00  & 0.42 &  0.65 & 0.97 \\ 
$h(z_2, z_3)$  &  SoftBart &  0.18 & 0.93 & 0.91 & 0.58 & 1.05 & 0.91  \\
               & BKMR        &  0.02 & $0.99$ & 0.96 & 0.42 & 0.65 & 0.96 \\
\toprule
\multicolumn{8}{c}{Case 2: $h(\bdz) = h^{(2)}(\boldsymbol z)$}\\
\midrule
$h(z_1, z_2)$  & SoftBart & 0.05 &  1.00 & 1.00  & 0.45 & 0.96 & 0.97  \\
               & BKMR       	       & 0.11 & 0.98  & 0.99  & 0.79 & 1.28 & 0.32 \\ 
$h(z_1, z_3)$  &  SoftBart & 0.01 & 1.00 &  1.00 &  0.63 & 1.55 & 0.91    \\  
               & BKMR        & 0.21 &  0.96 & 0.98  & 1.03 &  1.59 & 0.18  \\ 
$h(z_2, z_3)$  &  SoftBart & 0.03 & $0.99$ & $0.99$ & $0.63$ & 1.48 & 0.93  \\
               & BKMR        &  0.04 & 0.99 & 0.98 & 0.82 & 1.32 & 0.41 \\
\toprule
\multicolumn{8}{c}{Case 3: $h(\bdz) = h^{(3)}(\boldsymbol z)$}\\
\midrule
$h(z_1, z_2)$  & SoftBart &  0.13 & 0.97  & 0.99  & 2.12 & 6.23 & 0.64   \\
               & BKMR       	       & 0.02 & 1.00  & 0.99  & 0.99 &  1.61 & 0.00 \\ 
$h(z_1, z_3)$  &  SoftBart & 0.20 & 0.96  & 0.99  & 1.69  & 4.75 & 0.73   \\  
               & BKMR        & $-0.06$ &  1.01 & 0.99  & 0.81 &  1.27 & 0.23 \\ 
$h(z_2, z_3)$  &  SoftBart &  0.07 & 0.99 & 1.00 & 0.79 & 1.76 & 0.88  \\
               & BKMR        & 0.00 & 0.98 & 0.99 & 0.88 & 1.33 & 0.91 \\
\bottomrule
\end{tabular}
\end{table}

Tables \ref{tab:sim-main-n100} and \ref{tab:sim-int-n100} summarize the simulation results for estimating $h^{(1)}(z_2)$, $h^{(2)}(z_2)$, and $h^{(3)}(z_3)$ with the sample size $n = 100$. We first observe that even with a relatively small sample size and the high correlations among variables, both methods are successful in identifying the correct submodel. Second, by examining the $R^2$, which measures the goodness of fit between the posterior mean and the true values, both BKMR and SoftBart produce comparable results. Specifically, when the $R^2$ for BKMR is low, the $R^2$ for the SoftBart estimates is also low.

Nonetheless, there are several noticeable key differences between the two methods.
First and foremost, as we already mentioned in Figure \ref{fig:runtime}, SoftBart is computationally much faster in all simulation scenarios.
Second, we observed that the size of the credible bands for SoftBart is generally wider than those for BKMR when the true function is not constant. However, when the truth is constant (see $h^{(1)}(z_3)$ and $h^{(3)}(z_3)$) the credible bands for SoftBart are significantly narrower. This difference indicates that SoftBart is more adaptive in estimating functions with varying shapes, as the posterior distribution for BKMR must be Gaussian, whereas SoftBart does not need to be Gaussian, which is more flexible and less restrictive.
Third, examining the results for $h^{(3)}(z_1)$, where poor fitting is expected for BKMR, we observe that the credible bands of BKMR indeed exhibit much poorer coverage. This difference in coverage becomes even more pronounced when the sample size increases to 1000 (see Table \ref{tab:sim-main-n1000}). Notably, we find that $R^2$ alone is not a reliable indicator for evaluating the performance of an algorithm. For instance, despite the low coverage of the credible bands, the $R^2$ values for both methods remain quite high. Conversely, in cases where $R^2$ is low (e.g., $h^{(2)}(z_2)$), the coverage is not necessarily lower.
Last, we examine the two-way interaction effects. In Table \ref{tab:sim-int-n100}, we observe that, except for $h(z_1, z_3)$ and $h(z_2, z_3)$ in Case 1, where both methods achieve good coverage, SoftBart outperforms BKMR in coverage across nearly all scenarios. In Case 2, the BKMR method provides extremely low coverage for every function. In Case 3, the coverage of BKMR for $h(z_1, z_2)$ is only 0.14.

We now take a look at the results for $n = 1000$, which are presented in Tables \ref{tab:sim-main-n1000} and \ref{tab:sim-int-n1000} for main effects and interaction effects, respectively. We notice that as the sample size increases, both the $R^2$ and the coverage for SoftBart improve. However, the coverage of $h^{(3)}(z_1)$ for BKMR decreases to 0.18, and the coverage for the interaction effect $h^{(3)}(z_1, z_2)$ drops to 0. The SoftBart, on the other hand, provides much larger coverage probability than BKMR.
In sum, those simulation results indicate that the SoftBart method is more adept at estimating functions with various shapes than BKMR. In general, SoftBart provides better coverage, particularly for interaction effects and functions with varying smoothness levels.

Besides the aforementioned benefits, more importantly, the SoftBart method is computationally much faster than BKMR. Even with a relatively small sample size of $100$, we found that SoftBart is already three times faster than BKMR. The computational speed ratio between SoftBart and BKMR increases to 82 when the sample size reaches $1,000$. 

\subsection{Simulation study for 13 mixture components}
\label{sec:sim-m13}

In this section, we increase the number of variables from $M = 3$ to $M = 13$ and run simulations with a sample size of $n = 1000$. Due to computational resource constraints, we are unable to run BKMR for sample sizes larger than $n = 1000$.
Similar to the simulation results in Section \ref{sec:sim-m3}, we report the estimations for both the main and interaction effects in Table \ref{tab:sim-M13}. Estimates are provided only for the first three variables, as the remaining variables are constants and yield similar results as $h(z_3)$ in terms of $R^2$, coverage, and PIP.

Our findings in Table \ref{tab:sim-M13} are similar to those in the previous section. The $R^2$ of the posterior mean for SoftBart and BKMR are comparable, although in some cases, BKMR achieves a higher $R^2$ than SoftBart (see $h(z_2)$ and $h(z_2, z_3)$). In general, with a wider confidence band, the coverage of SoftBart is better than that of BKMR. One exception is the estimation of the interaction effect between $z_1$ and $z_2$, where the coverage for BKMR is 0.41, but for SoftBart, it is only 0.18. Despite BKMR having higher coverage in this case, both methods fall significantly below the nominal level of 0.95.
We did not observe a similar under-coverage phenomenon in Table \ref{tab:sim-int-n1000} for this interaction effect, likely because the correlation structure is inherently different, even though the true functions for the first three variables remain the same. We conducted an additional simulation exclusively for the SoftBart method with an increased sample size of $n = 10,000$. In this scenario, the coverage for this interaction effect increased to 0.95, highlighting SoftBart's potential to achieve nominal coverage levels when sufficient data are available.

\begin{table}[!h]
\centering
\caption{Simulation results of the SoftBart and BKMR methods ($n = 1000$). We report the intercept, slope, and $R^2$ for the regression of $h(\cdot)$ of $\hat h(\cdot)$, the lengths of the 95\% piecewise credible interval and the 95\% credible band, the coverage probability of the credible band, and the posterior inclusion probability (PIP).}
\label{tab:sim-M13}
\begin{tabular}{c c | c c c | c c c | c  }
\toprule
Main effect \&  & & \multicolumn{3}{c|}{Regression of $h(\cdot)$} & \multicolumn{3}{c|}{Uncertainty quantification ($\alpha$ = 0.95)} &   \\
Interaction effect & Method & Intercept & Slope & $R^2$ & len(C.I.) & len(C.B.)  & Coverage & PIP\\
\midrule
\multicolumn{9}{c}{Case 4: $h^{(4)}(\boldsymbol z)$}\\
\midrule
$h(z_1)$  & SoftBart &  $-0.37$ & 1.12 &  0.97 & 1.72 & 2.83 & 0.95 & 1.00  \\
               & BKMR       & 0.08 & 0.97 & 0.99  & 0.88 & 1.22 & 0.86 & 1.00 \\ 
$h(z_2)$  &  SoftBart & 0.23 &  0.92 &  0.56 &  1.73 & 2.91 & 0.95  & 1.00  \\  
               & BKMR        & 0.10 &  0.93 &  0.79 &  0.84 & 1.16 & 0.93  & 1.00   \\ 
$h(z_3)$  &  SoftBart & 4.32 & 0.01 & 0.04 & 0.28 & 0.24  & 1.00 & 0.00  \\
               & BKMR        & 2.00 & 0.00 & 0.02 & 0.30 & 0.30  & 1.00 & 0.00\\
$h(z_1, z_2)$  & SoftBart & 0.29 &  0.91 & 0.93  & 2.47 &  4.67 & 0.18  & --   \\
               & BKMR       	       &0.06 &  0.98 & 0.98  & 1.96 &  2.93 & 0.41  & --\\ 
$h(z_1, z_3)$  &  SoftBart & $-0.38$ &  1.12 &  0.97 &  1.75 & 2.86 & 0.98  & --  \\  
               & BKMR        & 0.10 &  0.97 &  0.98 &  0.92 & 1.27 & 0.85  & --  \\ 
$h(z_2, z_3)$  &  SoftBart & 0.16 & 0.95 & 0.57 & 1.77 & 2.94 & 0.95 & -- \\
               & BKMR        &   0.12 & $0.91$ & $0.78$ & $0.88$ & 1.22 & 0.92 & -- \\
\midrule                                 \\[-6pt]
\multicolumn{2}{c}{Avg. running time} & \multicolumn{6}{c}{SoftBart: 157.23s, \quad BKMR: 14158.24s} & \\[2pt]               
\bottomrule
\end{tabular}
\end{table}

\section{The NHS data study using SoftBart }
\label{sec:real}

In this section, we apply the SoftBart method to analyze the NHS data \citep{xu24}. This dataset contains 10,110 participants who underwent the forth cognitive assessment between 1995 and 2008. During the assessment, each individual completed a series of cognitive tests, including a telephone interview of cognitive status (TICS) \citep{sterr02}, delayed recall from the TICS 10-word list, immediate and delayed recall from the East Boston Memory Test (EBMT), a category fluency task (animal naming test), and the digit span backward test \citep{baddeley91}. Because these tests operate on different scales, we follow \citep{xu24} to standardize the results by converting them into z-scores and computed a global composite score by averaging these values to represent overall cognitive function. 
To obtain the environmental exposure, we calculated a 12-month moving average of PM2.5 and its constituents using data collected in the year preceding each participant’s fourth cognitive assessment. This estimate was treated as a proxy for actual exposure levels. Detailed methodological descriptions, including participant selection, cognitive testing procedures, and exposure assessment methods, can be found in previous reports \citep{weuve12}. In addition, we also obtain several covariates as confounding variables for each participate including age, body mass index (BMI), smoking history, and educational background.

The cognitive function is calculated using z-score to create a global composite score as the average of all z-scores of those cognitive tests. Monthly data of 12 PM2.5 constituents, including Br, Ca, Cu, Fe, Mn, Ni, S, Se, Si, Ti, V, and Zn, are recorded using concentrations measured at the monitors from the U.S. EPA database\footnote{The database can be found in the Air Quality System (\url{https://aqs.epa.gov/aqsweb/airdata/download_files.html})}. In addition, BMI, smoking status, and education level of each participant are collected and treated as confounding variables. 
Note that we did not perform the analysis for BKMR because the sample size is too large for it to handle. In contrast, the SoftBart method is computationally efficient and can complete the MCMC sampling with 20,000 iterations within a few hours.
We take the log transformation of 12 PM2.5 constituents before fitting into the BKMR and SoftBart methods.

\begin{figure}
\centering
\includegraphics[width=0.9\textwidth]{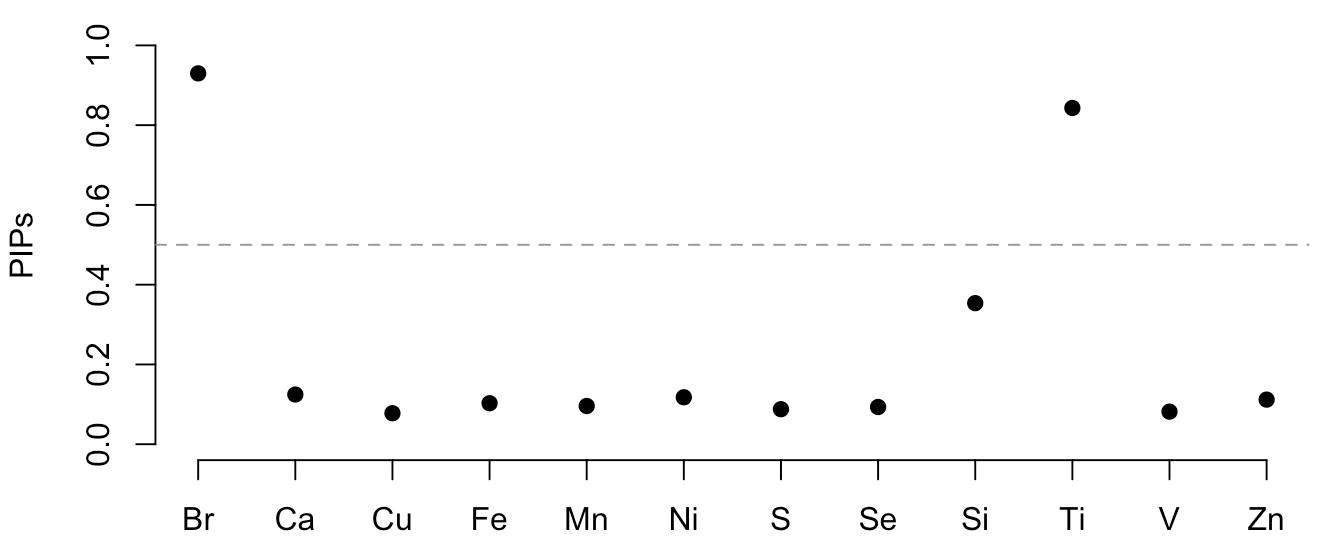}
\vspace{-0.4cm}
\caption{PIPs of PM2.5 constituents. The gray dashed line indicating the probability at 0.5.}
\label{fig:pips}
\end{figure}

We first compute the PIP for each of the 12 PM2.5 constituents (see Figure \ref{fig:pips}). The gray line indicates the threshold at 0.5. There are two constituents, Br and Ti, exceed the threshold, with PIPs of 0.93 and 0.84 respectively. The PIP of Si is 0.35, making it the third largest, while the PIPs of the remaining constituents are all smaller than 0.13. 

%\citet{xu24} studied the same dataset as ours and also identified Br and Ti as constituents significantly correlated with cognitive function. In addition to these two, they also identified Mn and Ni as significant constituents, which were not selected in our results. However, our method differs fundamentally from the approach of \citet{xu24} in two key aspects. First, their approach analyzes one constituent at a time while treating the remaining 11 constituents as confounding variables, whereas our method includes all constituents simultaneously in the model. Second, their approach uses linear regression  controlling confounding variables. In contrast, our method is fully nonparametric for modeling $h(\cdot)$, allowing for nonlinear modeling of main and interaction effects, and even higher-order interactions. 

\begin{figure}[!h]
\centering
\includegraphics[width=0.46\textwidth]{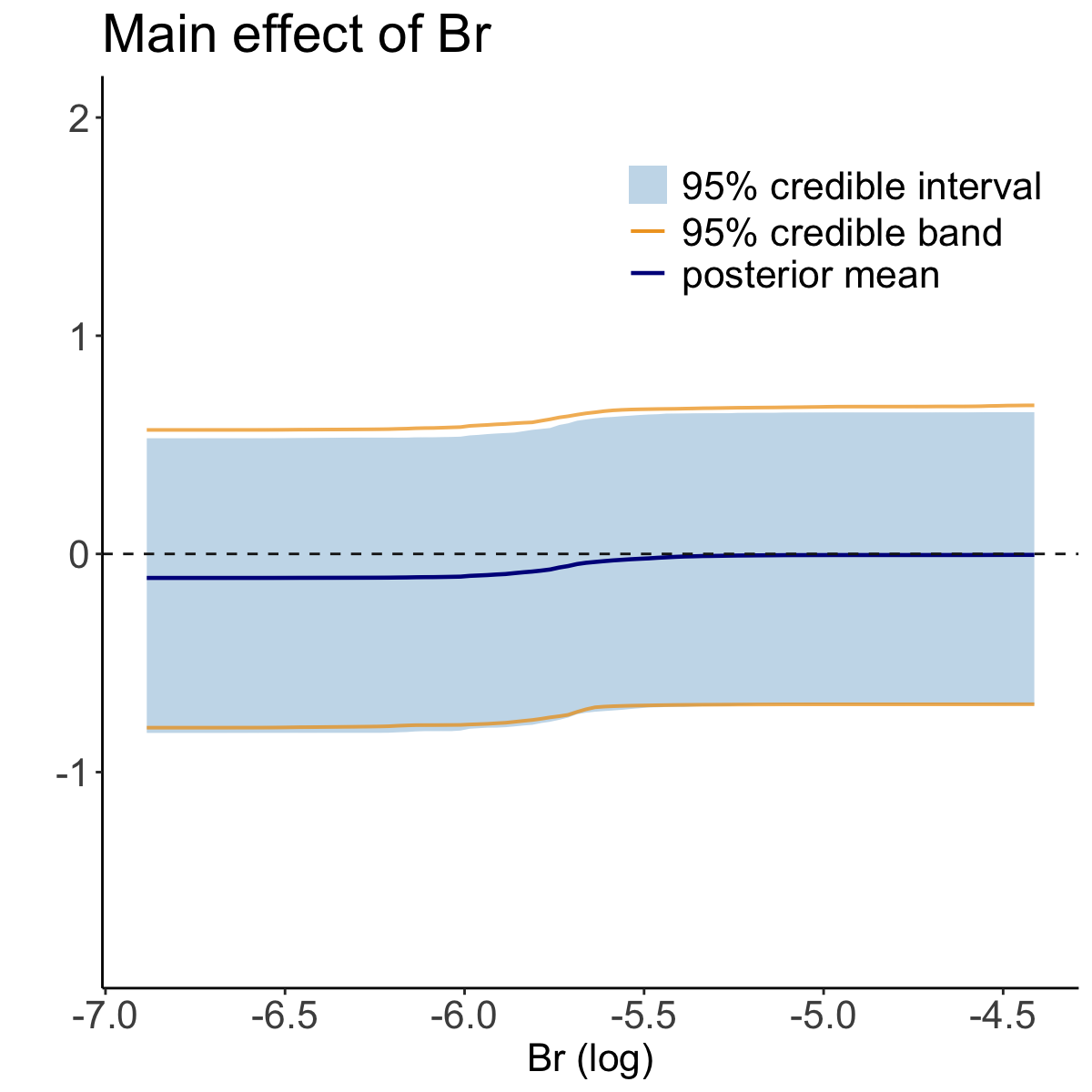}%
\quad \includegraphics[width=0.46\textwidth]{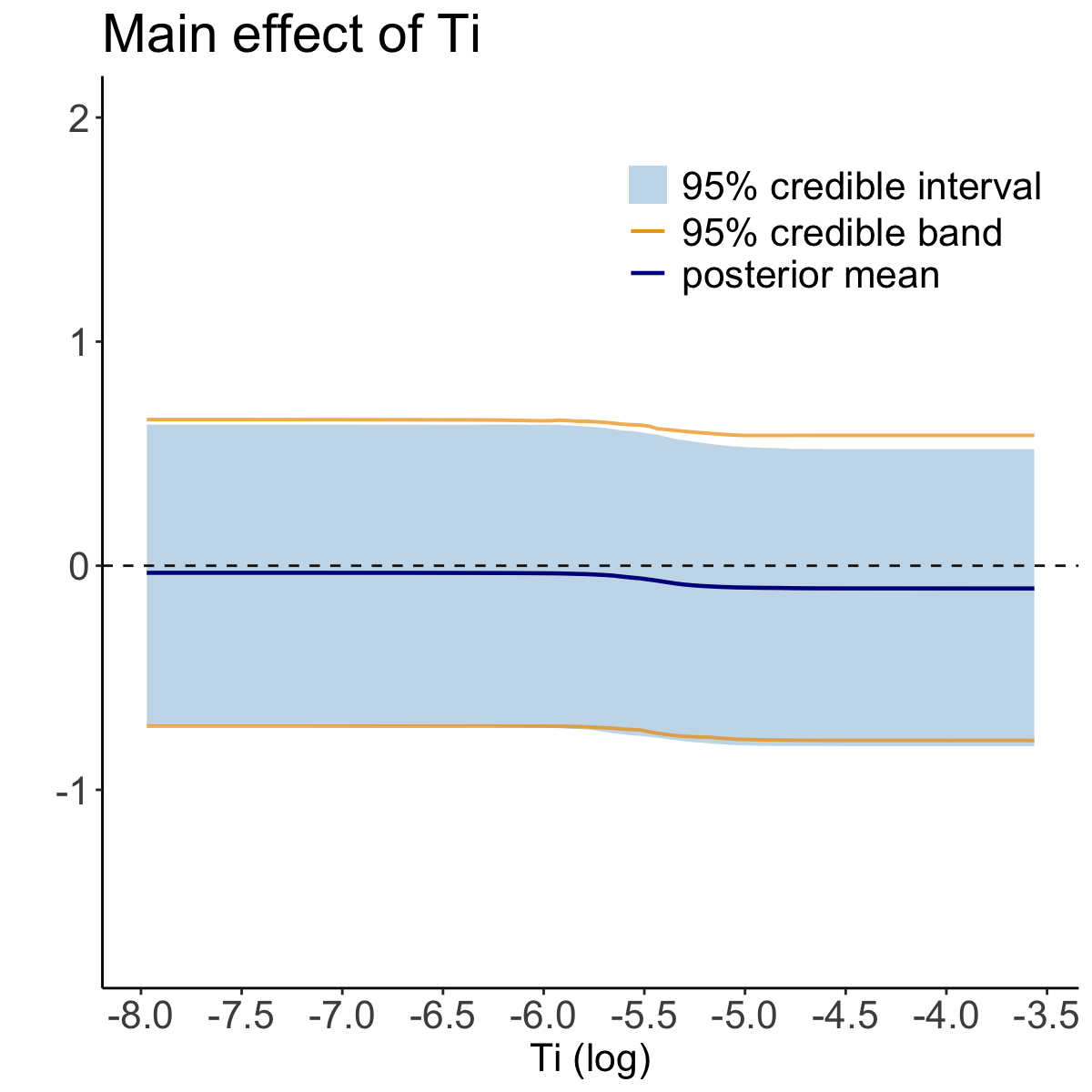}\\[2mm]
\includegraphics[width=0.46\textwidth]{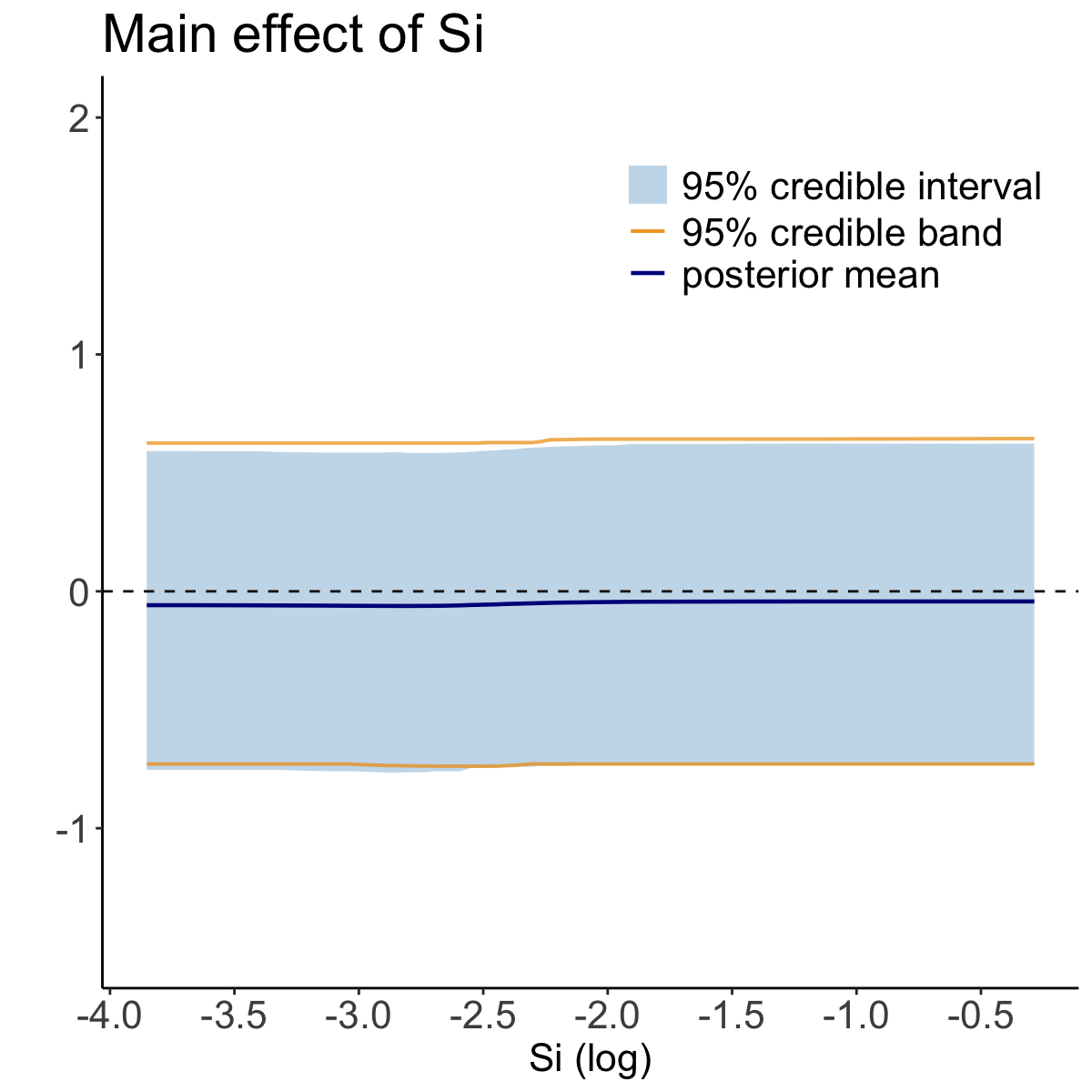}
%\vspace{-0.2cm}
\caption{The estimated main effects of Br (left), Ti (middle), and Si (right). In each subplot, the overall mean is removed. The dark blue line is the posterior mean, the blue region indicates 95\% pointwise credible intervals and the two orange lines are the upper and lower bounds for 95\% credible band.}
\label{fig:main-effect-br-ti}
\end{figure}

In Figure \ref{fig:main-effect-br-ti}, we plot the estimated main effects of Br, Ti, and Si as a function of their concentrations. The 95\% pointwise credible intervals and credit bands are shown for an illustrative purpose; if using the 95\% credible intervals and bands, which are wider, we would have to change the scale in the y-axis, and as a result,  the curvature of the main effect would not be displayed as clearly in the same panel. 
%Note that the overall mean of $\hat h(\bdz)$ is removed before calculating the main effects.  
%The posterior means for the main effects of Br and Ti are both negative. The posterior mean for the main effect of Br suggests an interesting phenomenon that the function is monotone and increases as its concentration rises. It then staying flat around zero after the concentration exceeds $e^{-5.3}$. The slope of this function is not constant; it reaches the maximum value between $e^{-5.3}$ and $e^{-5.7}$.In contrast, Ti exhibits a monotonic decreasing relationship with cognitive function as its concentration increases, with the steepest decline occurring between the concentration values $e^{-5.7}$ and $e^{-5.1}$. The posterior mean for the main effect of Si remains nearly constant, suggesting a minimal effect on cognitive function. This aligns with its exclusion from the model, as indicated by the low PIP value shown in Figure \ref{fig:pips}.

\begin{figure}[!h]
\centering
\includegraphics[width=0.48\textwidth]{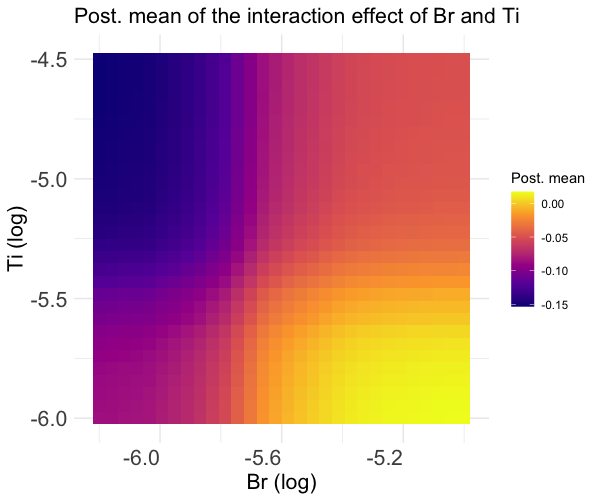}%
\quad \includegraphics[width=0.48\textwidth]{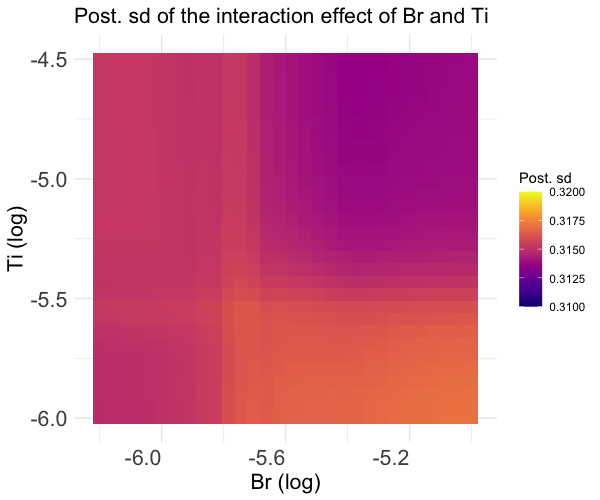}
\vspace{-0.4cm}
\caption{Posterior mean (left) and sd (right) of the interaction effect of Br and Ti.}
\label{fig:int-effect-brti}
\end{figure}

\begin{figure}[!h]
\centering
\includegraphics[width=0.5\textwidth]{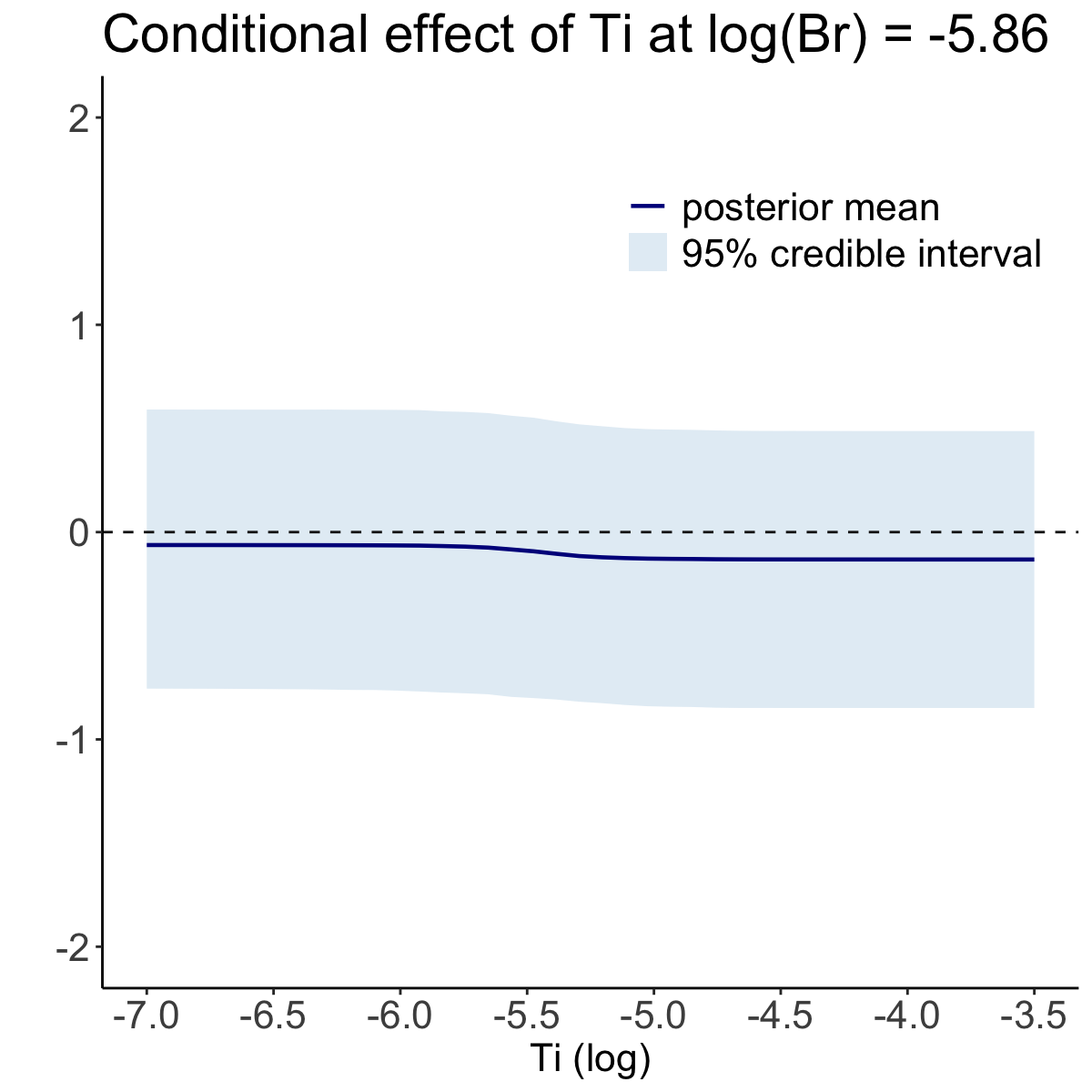}
\vspace{-0.4cm}
\caption{The conditional effect of Ti at Br $= e^{-5.83}$. The dark blue line is the posterior mean and the blue region indicates 95\% pointwise credible intervals.}
\label{fig:cond-tibr}
\end{figure}

We further plot their interaction effect in Figure \ref{fig:int-effect-brti}. The left plot in the figure shows the posterior mean of the interaction effect over a grid of concentration values for Br and Ti, while the plot on the right-hand side displays the corresponding posterior standard deviation.
The posterior mean indicates that the combination of Ti and Br may have the most detrimental effect on health when the concentration value of Br is less than about $e^{-5.6}=0.0037$ and that of Ti is larger than $e^{-5.5}=0.0041$. 
This effect gradually decreases as the concentration value of Br increases and that of Ti decreases. On the other hand, the posterior standard deviations across various values of Br and Ti remain stable, with values between 0.31 and 0.32.
We further plot the conditional effect of Ti when Br $= e^{-5.83} = 0.0029$ in Figure \ref{fig:cond-tibr}. The figure shows that Ti exhibits a slowly decreasing function. 
%By investigating different values of Br, we observe a similar relationship for the conditional effect of Ti.

\section{Discussion}
\label{sec:disc}

In summary, this paper introduces the SoftBart method to estimate the relationship between health outcomes and multiple correlated multipollutant constituents. This method is flexible and computationally efficiency, as emphasized in Figure \ref{fig:runtime} and in the simulation study (see the running time in Tables \ref{tab:sim-main-n100} \& \ref{tab:sim-main-n1000}), making it particularly suitable for analyzing large-scale datasets. 
We have compared it with the commonly used method, BKMR, and our simulation results suggest that the SoftBart method excels in estimating interaction effects among highly correlated variables and modeling complex functions with varying degrees of smoothness.
Finally, to illustrate the softBart method, we have applied it to the NHS dataset to evaluate the relationship between PM2.5 multipollutants and cognitive function. 
In our simulation and real data study, we choose the number of trees to be a fixed value 30. 
In our simulation and real data studies, we found that selecting any value between 10 and 40 for the number of trees did not substantially affect the results.
One could use cross-validation to select the optimal number of trees. Alternatively, one could place a prior on the number of trees; however, the computational burden would significantly increase, as the Reversible-jump Markov Chain Monte Carlo algorithm needs to be adopted to sample the posterior of the number of trees. Hence, we do not recommend choosing this strategy.

The current work opens new avenues for modeling multipollutant data using the Bayesian ensemble learning approach. Since health outcomes in environmental studies are frequently survival data. We play to extend the SoftBart approach to incorporate group selection features within the Cox model framework or using the probit regression model for survival data.

%First and foremost, a topic for future research  is to develop a extension of the SoftBart method specifically designed to address measurement error issues. 
%Second, since multipollutant constituents are highly correlated, and it has been shown that grouping these variables before performing selection can improve both interpretability and predictive performance. Additionally, health outcomes in environmental studies are frequently survival data. (link to probit) An important direction is to extend the SoftBart approach to incorporate group selection features within the Cox model framework for survival data.
%Finally, environmental exposures and health outcomes often exhibit nonlinear and time-varying relationships. The third direction is to improve the current SoftBart method to model these relationships effectively, enhancing its applicability to data analysis in such scenarios.

\section*{Acknowledgement}

This work was supported by the NIH grant R01ES026246.

%\nocite{*}% Show all bib entries - both cited and uncited; comment this line to view only cited bib entries;

%%%%%%%%%%%%%%%%%%%%%%%%%%%%%%%%%%%%%%%%
%%%%%%%%%%%%%%%% Bibliography %%%%%%%%%%%%%%%%%
%%%%%%%%%%%%%%%%%%%%%%%%%%%%%%%%%%%%%%%
\bibliographystyle{chicago}
\bibliography{citation.bib}

\end{document}